# Einstein-Podolsky-Rosen Entanglement between Separated Atomic Ensembles

Wei Zhang[*], Ming-Xin Dong[*], Dong-Sheng Ding[†], Shuai Shi, Kai Wang, Shi-Long Liu, Zhi-Yuan Zhou, Guang-Can Guo, Bao-Sen Shi[#]

[1]Key Laboratory of Quantum Information, CAS, University of Science and Technology of China, Hefei, Anhui, 230026, China

[2]Synergetic Innovation Center of Quantum Information & Quantum Physics, University of Science and Technology of China, Hefei, Anhui, 230026, China

[*]*These authors contributed equally to this work.*

*Corresponding author:* [†]*dds@ustc.edu.cn*

[#]*drshi@ustc.edu.cn*

The Einstein-Podolsky-Rosen (EPR) entanglement is of special importance not only for fundamental research in quantum mechanics, but also for quantum information processing. Establishing EPR entanglement between two memory systems, such as atomic ensembles, nitrogen-vacancy centers, and rare-earth-ion-doped solids is promising for realizing spatial-multimode based quantum communication, quantum computation, and quantum imaging. To date, there have been few reports on realizing EPR entanglement between physical systems in true position and momentum. Here we describe successfully establishing such entanglement between two separated atomic ensembles by using quantum storage. We clearly prove the existence of EPR entanglement in position and momentum between two memories by demonstrating the satisfying of the inseparability criterion with the aid of quantum ghost-imaging and ghost-interference experiments.

The original Einstein-Podolsky-Rosen (EPR) paradox involving a pair of particles that were perfectly correlated in position and anti-correlated in momentum was introduced in 1935 [1]. Such a state violates the Heisenberg inequality for the product of conditional variances of positions and variances of momenta, which suggested the failure of local realism. Since then, EPR paradox has attracted wide attention because of its fundamental role in identifying the nonlocality. Until 2002, another inequality was derived as inseparability criterion [2], which is weaker than paradox criterion, to determine whether a state is inseparable, i.e. EPR entangled. In past several decades,

EPR entanglement was demonstrated in continuous variable (CV) systems, which has many promising applications in quantum information field including quantum imaging and quantum metrology [3–5], quantum computation [6], and quantum communication [7].

As a featured example of CV entanglement, EPR entanglement in position and momentum has remained topical for many researchers. The initial experimental demonstration of entanglement was conducted with polarization-correlated photons for verifying non-local features of quantum mechanics [8]. Later, this momentum-position-like entanglement was widely explored in a variety of systems including bulk crystals [9–11], squeezed fields [12, 13], and atomic ensembles [14]. The two photons generated through spontaneously parametric down conversion in a nonlinear crystal were identified as being entangled in momentum and position [10], which provided a direct way to distinguish quantum entanglement from classical correlations. In addition, long-lived entanglement between two macroscopic atomic ensembles was obtained through a non-local Bell measurement, which is suitable for atomic teleportation [14]. Very recently, light-atom EPR paradox in true position and momentum bases is reported in a hot atomic ensemble, where a 12-dimensional entangled state is prepared [15].

For quantum communications, EPR entanglement is very useful in realizing CV protocols including unconditional quantum teleportation [16, 17], quantum key distribution [18], and high-dimensional quantum communication [15, 19, 20]. Quantum communication protocols require entanglement to be distributed or transferred synchronously between different nodes [21, 22]. Therefore, realizing EPR entanglement between memories, such as atomic ensembles, nitrogen-vacancy centers or rare-earth-ion-doped solids, is indispensable. Although certain progress has been made in this direction including realizing quantum memory for entangled CV states [23], tunable delay for EPR entanglement in quadrature amplitudes [24], EPR entanglement in position and momentum between light and spin waves [15] and even multiple degrees-of-freedom entanglement between two quantum memories [25], the EPR entanglement of atomic spin waves between memories in true position and momentum bases has never been reported, and its experimental realization is still lacking.

Here we report on the first experimental realization of EPR entanglement of atomic spin waves in position and momentum between separated atomic ensembles by using quantum storage, in which entanglement is verified through photonic ghost-imaging and ghost-interference experiments. In our experiments, the EPR entanglement between a photon and a spin wave is established firstly through spontaneous Raman scattering (SRS) in one cold atomic ensemble. This single photon is then delivered and stored in another cold atomic ensemble as a spin wave using the Raman protocol [26], thus established the entanglement between two atomic ensembles. Finally, we retrieved both spin waves to single photons and check their entanglement by performing quantum imaging and quantum ghost experiments. We clearly prove the existence of EPR entanglement before and after storage by demonstrating that the inseparability criterion is satisfied. Our experimental results demonstrate the realization of EPR entanglement of atomic spin waves in true position and momentum entities between two atomic ensembles for the first time.

In this experiment, we use the quantum ghost interference-imaging effects to distinguishing quantum entanglement in momentum and position space [10, 27, 28]. Simply speaking, ghost interference and ghost imaging experiment are performed with a pair of photons, where an obstacle is placed in one photon's path and then this photon is detected by a detector with no spatial resolution, the other photon is detected by a detector with spatial resolution like Charge Coupled Device or a scanning detector, the coincidence of the photon pair would show the imaging and interference phenomena. We then can calculate the position-momentum correlation from the observed ghost imaging and ghost interference.

**Result**

The medium used here to generate EPR entanglement is an optically thick ensemble of $^{85}$Rb atoms trapped in a two-dimensional magneto-optical trap (MOT) [29]. Fig.1 is simplified setup. Our system works periodically with a cycle time of 10 ms including 8.7-ms trapping time and an initial state preparation time, and 1.3-ms operation time containing 2600 cycles (tunable) with a cycle time of 500 ns. In each cycle, Pump-1, Pump-2, and Coupling light are pulsed by acousto-optic modulators. Pump 1 and Pump 2 are Gaussian beams with a waist of $\omega_0 = 1.1$ mm. The optical depth of the atomic ensembles in MOT A and MOT B are about 40 and 50, respectively. A metal bar with a block width of $\omega_b = 1.04$ mm inserted in the path of Signal 2 provides an effective double slit with the aid of fiber collimator FC 2, which can output a

Gaussian collimated beam with waist $\omega_0$.

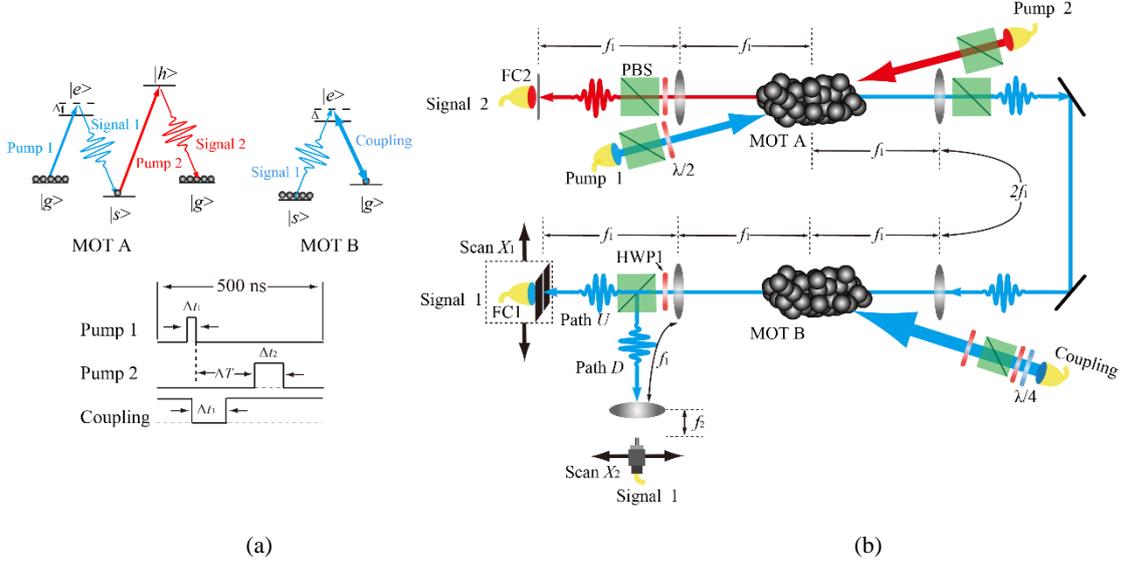

(a)                                                        (b)

FIG. 1. (a) Energy diagram and timing sequence. Pumps 1 and 2 are pulses with durations $\Delta t_1$=30 ns and $\Delta t_2$=200 ns. Storage time set by $\Delta T$ for the spin wave in MOT A is 200 ns and $\Delta t_3$=100 ns for the spin wave in MOT B. The blue detuning of Pump 1 is set to +70 MHz; the related energy levels are $|s\rangle$=$|5S_{1/2}$, F=2$\rangle$, $|g\rangle$=$|5S_{1/2}$, F=3$\rangle$, $|e\rangle$=$|5P_{1/2}$, F=3$\rangle$, and $|h\rangle$=$|5P_{3/2}$, F=3$\rangle$. (b) Experimental setup. PBS: polarizing beam splitter; $\lambda$/2: half-wave plate (HWP); $\lambda$/4: quarter-wave plate; FC: fiber collimator enclosed with a lens with a focal length of $f_c$=11.07 mm; $f_1$ =500 mm and $f_2$ =32 mm. Slit width is $\omega_d$=0.4 mm. Collimated Pump 1 is incident on the atomic ensemble in MOT A, making a 2° angle with the direction of Signal 1, and Pump 2 counter-propagates with Pump 1 through the atomic ensemble. The Coupling light is also incident on the atomic ensemble in MOT B with an angle of 3° with respect to the path of the Signal-1 photon. The Signal-1 photon is collected through two different paths (*U* or *D*) by tuning the optical axis of HWP1 before PBS: Path *U* is used for ghost imaging and path *D* is used for ghost interference. The powers of Pump 1, Pump 2, and Coupling light are 0.1 mW, 4 mW, and 30 mW, respectively.

In the simplified experimental setup (Fig. 1), a Signal-1 photon is directly generated through SRS after Pump-1 light illuminates the atomic ensemble in MOT A. Because momentum is conserved in the SRS process and the initial system has zero momentum, the Signal-1 photon and the spin wave in MOT A are anti-correlated in momentum. This hybrid state can be denoted as [15, 28],

$$|\psi_{\text{A-S1}}\rangle = \int dp_A dp_{S1} \overline{\phi}_1(p_A, p_{S1}) a_A^\dagger a_{S1}^\dagger |00\rangle, \qquad (1)$$

where $p$ is transverse momentum, $a^\dagger$ is creation operator, subscript A represents atomic spin wave in MOT A and subscript S1 represents Signal-1 photon. We define that $|\phi_A\rangle = a_A^\dagger |0\rangle$ which represents an atomic spin wave generated in MOT A. The wave function of $|\psi_{\text{A-S1}}\rangle$ in the transverse momentum space is defined with Gaussian shape [11, 15, 28],

$$\overline{\phi}_1(p_A, p_{S1}) = \frac{\sigma_+ \sigma_-}{\pi} \exp(-\sigma_+^2 \frac{|p_A + p_{S1}|^2}{4\hbar^2} - \sigma_-^2 \frac{|p_A - p_{S1}|^2}{4\hbar^2}) \ . \tag{2}$$

By Fourier-transforming the wave function in transverse momentum space, we can get the wave function in transverse position space [11, 15],

$$\phi_1(r_A, r_{S1}) = \frac{1}{\pi \sigma_+ \sigma_-} \exp(-\frac{|r_A - r_{S1}|^2}{4\sigma_-^2} - \frac{|r_A + r_{S1}|^2}{4\sigma_+^2}) , \tag{3}$$

where variable $r$ represents the corresponding position. Here $\Delta^2 |r_A - r_{S1}| = \sigma_-^2$ and $\Delta^2 |p_A + p_{S1}| = \frac{1}{\sigma_+^2}$, where $\Delta$ signifies the standard deviation. We can see that the spin wave and Signal-1 photon are also position correlated. By this way, we can prepare the position-momentum correlation between an atomic spin wave in MOT A and a Signal-1 photon.

We next deliver the Signal-1 photon to the second atomic ensemble in MOT B for storage. With the shutting down of the Coupling light adiabatically by using an acousto-optic modulator, the Signal-1 photon is stored in MOT B using the Raman protocol [26]. Under phase matching condition, Signal-1 photon with transverse momentum $p_{S1}$ is converted to an atomic spin wave in MOT B with transverse momentum $p_B$, thus establishing the momentum correlation of atomic spin waves between two atomic ensembles. We can also get their correlation in position space through applying Fourier-transforming method. The obtained momentum-position correlation can be denoted as,

$$\begin{aligned} |\psi_{A\text{-}B}\rangle &= \int dp_A dp_B \overline{\phi}_2(p_A, p_B) |\phi_A \phi_B\rangle \\ &= \int dr_A dr_B \phi_2(r_A, r_B) |\phi_A \phi_B\rangle \end{aligned}, \tag{4}$$

where $|\phi_B\rangle$ is atomic spin state in MOT B, $\phi_2$ is the Fourier transformation of $\overline{\phi}_2$. $\phi_2$ and $\overline{\phi}_2$ are wave function with Gaussian shape.

$$\overline{\phi}_2(p_A, p_B) = \frac{\sigma'_+ \sigma'_-}{\pi} \exp(-\sigma'^2_+ \frac{|p_A + p_B|^2}{4\hbar^2} - \sigma'^2_- \frac{|p_A - p_B|^2}{4\hbar^2})$$

$$\phi_2(r_A, r_B) = \frac{1}{\pi \sigma'_+ \sigma'_-} \exp(-\frac{|r_A - r_B|^2}{4\sigma'^2_-} - \frac{|r_A + r_B|^2}{4\sigma'^2_+}) \tag{5}$$

Here, the $\Delta^2 |r_A - r_B| = \sigma'^2_-$ and $\Delta^2 |p_A + p_B| = \frac{1}{\sigma'^2_+}$. Through this method, we can prepare the position-momentum correlation between atomic spin waves in separated atomic ensembles.

For checking whether these momentum-position correlations belongs to the EPR

entanglement, one has to measure variances of positions and variances of momenta. For an ideal EPR entangled pair (particle *a* and particle *b*), $\Delta|r_a - r_b|=0$ & $\Delta|p_a + p_b|=0$, which is not accessible due to experimental imperfections. Then, a more practical way is to check whether they satisfy the original EPR paradox criterion. In transverse position and momentum bases, the EPR paradox occurs when [30–31]:

$$\langle \Delta^2 |r_a - r_b| \rangle \langle \Delta^2 |p_a + p_b| \rangle < \hbar^2/4 . \tag{6}$$

This criterion is also called EPR steering criterion [32], which relates to another very interesting research direction. Here, even if the product is bigger than $\hbar^2/4$, the bipartite state may still be entangled when the inseparability criterion is satisfied [2, 15, 28], which is given by

$$\langle \Delta^2 |r_a - r_b| \rangle \langle \Delta^2 |p_a + p_b| \rangle < \hbar^2 . \tag{7}$$

The inseparability criterion also has another form for general continuous variable systems [33, 34]: $V_r^- + V_p^+ < 2$, where $V_r^- / V_p^+$ represents the unified variance of position/momentum difference/sum. In general, the position-momentum correlation can be classified into different regimes depending on two criteria, EPR paradox criterion and inseparability criterion. When inseparability criterion is satisfied, this state can be called EPR entangled state, while EPR paradox, which requires a certain amount of entanglement, is a stronger EPR entangled state. Moreover, it is classical correlation when inseparability criterion is violated.

To verify the EPR entanglement between a spin wave and a Signal-1 photon or between atomic spin waves in separated ensembles, we retrieved the spin waves to single photons to evaluate the corresponding uncertainties, where the transverse momentum (position) of atomic spin waves can convert to transverse momentum (position) of photons under the phase matching condition. Experimentally, after storing the spin wave for 100-ns in MOT B and the spin wave in MOT A for 200-ns, we converted both spin waves to single photons by turning on the Coupling light and Pump-2 light, respectively. By this way, we can check the correlation between atomic spin waves through checking the correlation of retrieved photons via quantum-interference and quantum-imaging experiments [10, 28].

The ghost-imaging experiment is performed by adjusting the optical axis of HWP1 to the vertical, and therefore Signal 1 is collected through path *U* into FC 1. The FC 1 and a slit with a width of $\omega_d=0.4$ mm are stabilized on an electrical translation stage (ETS) with a precision of 0.078 µm. We scan the ETS in incremental steps of 0.1/0.2 mm recording the coincidence rates (coincidence resolution: 0.8 ns) between Signal 1 and Signal 2 against the transverse position $X_1$, and thus obtain the ghost-imaging data.

Then, we adjust the optical axis of HWP1 mentioned above to 45° with respect to the vertical to perform the ghost-interference experiment. In this case, the Signal 1 photon is conducted through path $D$ and a lens ($f_2$=32 mm) into another fiber whose head (inside diameter: ~5 µm) is localized on another ETS. Experimentally, we record the coincidence rate against the change in the transverse position $X_2$ in incremental steps of 2 µm.

We want to emphasize that both ghost-imaging and ghost-interference experiments must be performed under precise configurations. For the former, the metal bar should be imaged on the vertical slit in path $U$ through two sequential 4F imaging systems; for the latter, the metal bar is imaged on the lens in path $D$ and the fiber head is localized on its focal plane. Moreover, the two atomic ensembles are 4F-imaged with each other. With this exquisite arrangement, we can perform the experiments to demonstrate EPR entanglement.

For the EPR entanglement between a spin wave in MOT A and a Signal-1 photon, we convert the spin wave as the Signal-2 photon after 200-ns storage and perform quantum ghost-imaging and ghost-interference experiments, respectively, simultaneously shutting down MOT B. The experimental results are shown in Fig. 2 (a) and (b). For the EPR entanglement between two spin waves in separated atomic ensembles, we convert both spin waves to photons after a 100-ns storage for the spin wave in MOT B and a 200-ns storage for the spin wave in MOT A. Fig. 2 (c) and (d) show the experimental results. Please see Methods for more experimental details.

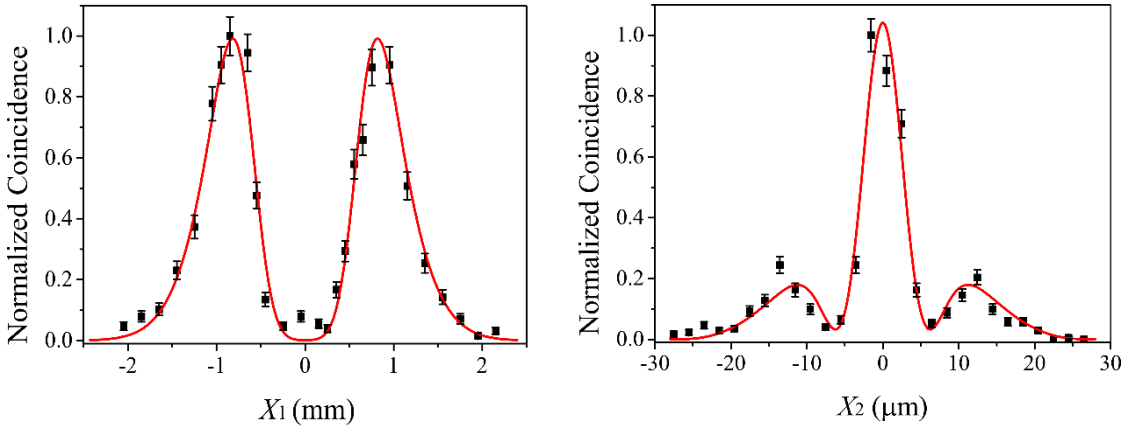

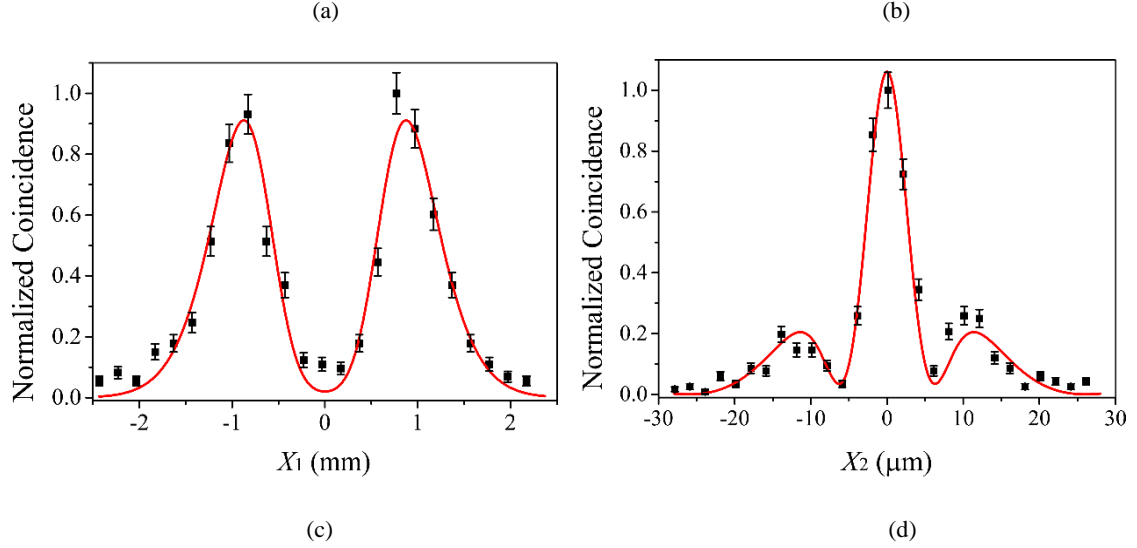

FIG. 2 Experimental results for ghost imaging (a)/(c) and ghost interference (b)/(d) for EPR entanglement between the spin wave in MOT A and a Signal-1 photon/the spin wave in MOT B. The black dots are experimental data without subtracting any noise; red curves are theoretical fitting curves. The error bars are estimated from Poisson statistics and represent ±s.d.

Although one may observe ghost imaging and ghost interference using classical position-correlated and momentum-correlated photons, it is impossible to observe both with high contrast and high resolution simultaneously [27, 35]. Here we clearly see both observed here with high contract and high resolution which means the correlation must be non-classical. By comparing the resulting fitting curves with the theoretical curves [10], $\Delta p_{x+}/\Delta x_{-}$, which represents $\Delta(p_{xS1}+p_{xS2})/\Delta(x_{S1}-x_{S2})$, can be obtained [See Methods], the related values are listed in Table 1.

TABLE 1 Uncertainties for EPR entanglement

| EPR entanglement | $(\Delta p_{x+})^2 (\hbar^2/mm^2)$ | $(\Delta x_{-})^2 (mm^2)$ | $(\Delta p_{x+})^2 (\Delta x_{-})^2 (\hbar^2)$ |
|---|---|---|---|
| Spin wave – S1 photon | 0.807±0.163 | 0.230±0.021 | 0.186±0.041 |
| Spin wave – spin wave | 1.439±0.214 | 0.332±0.026 | 0.478±0.080 |

From the table, the measured results for position-momentum state between Signal-1 photon and a spin wave in MOT A clearly show EPR paradox, which is a stronger entangled state. From the measured data for entanglement between spin waves in MOT A and B, we can see this product satisfies the inseparability criterion, indeed demonstrates the existence of EPR entanglement between spin waves in separated atomic ensembles. Further, we observe a degradation for the EPR entanglement after a more step of storage through comparing data shown in the table. This may come from two aspects, one is the slight increment in noise level shown Fig. 2 and experimental imperfections, the other is the atomic decoherence of spin wave in MOT B which will make a decrease for the level of entanglement [15]. Definitely, if we increase the storage time of EPR entanglement between spin waves, we would observe a further decrease of entanglement level and even to classic regime due to atomic decoherence and lower signal-to-noise ratio caused by lower memory efficiency. In addition, here we only measure the EPR entanglement in $x/p_x$ space, however the symmetry of the system will guarantee a near same result for EPR entanglement in $y/p_y$ and $r/p_r$ space.

**Discussion**

In this experiment, we would get ~10 photon-photon coincidence counts per second before storing Signal-1 photon in MOT B, and the 100-ns storage efficiency in MOT B is estimated to be ~25% in detecting level with no metal bar before FC 2 and no slit in path $U$ before FC 1. The retrieved signal has a Signal to Noise Ratio of ~30. Because our magnetic field for trapping can't be shut down completely within 1.3 ms due to the big value of inductance, therefore the memory time is rather limited to 1.4 μs as discussed before in previous work [36]. In general, memory time can be improved by compensating the magnetic field or by using magnetic field-insensitive states, reducing atomic motion by using optical lattice and using dynamic decoupling method [37–41]. Memory efficiency can be increased by using waveform modulation [42].

In summary, we experimentally realized EPR entanglement between two atomic ensembles. To our knowledge, this is the first report of EPR entanglement between memory systems in true position and momentum. This setup for EPR entanglement is very promising for realizing spatially-multiplexed quantum information processing, high-dimensional quantum repeaters, quantum imaging, and quantum metrology. In particular, combining the EPR entanglement with entanglement in orbital angular momentum degrees of freedom [36, 43–45], or in polarization degrees of freedom [26], multi-pixel entanglement [46, 47] and super-high-dimensional hyperentanglement [48] can be achieved, which could help in building a versatile platform to realize quantum computing protocols and sophisticated quantum networks.

# Methods

## Details for coupling, filtering and data collecting

The coupling efficiency for Signal 1 is 75% for Signal 1 in path $U$ in the case of no metal bar in the front of FC2 and no vertical slid in front of FC1. The coupling efficiency for Signal 1 in path $D$ with no metal bar before FC2 is ~38%, this low efficiency is due to the mismatch of numerical aperture for FC2 (focal length $f_c$=11.07 mm) and lens ($f_2$=32mm).

Signal-1 photons are filtered using three homemade cavities (including temperature control) with 45% transmittance with 70dB isolation. Signal-2 photons are filtered using two homemade cavities with 65% transmittance with 40 dB isolation. The efficiency of single-photon detector for detecting Signal 1 and Signal 2 is ~50%.

For EPR entanglement between spin wave and Signal-1 photon, data point in Fig. 2 is obtained through accumulating coincidences in 1000 s (max counts: 252 at $X_1$~-0.8 mm) for ghost imaging and in 200 s (max counts: 344 at $X_2$=-1.5 μm) for ghost interference. For EPR entanglement after storage of Signal-1 photon as shown in Fig. 3, accumulating time becomes 3000 s (max counts: 219 at $X_1$~0.8 mm) for ghost image and 500 s (max counts: 290 at $X_2$=0 μm) for ghost interference. That the data-collecting time for ghost imaging experiment is rather long is due to the low coupling efficiency caused by the existence of slit before FC 1.

## Theoretical calculations

In this experiment, the metal bar with a block width of $\omega_b$=1.04 mm and the fiber collimator FC 2 which can output a collimated Gaussian beam with waist $\omega_0$=1.1 mm, constitute the effective double slit. The transfer function can be denoted as $\Gamma_x = e^{-x^2/\omega_0^2}$ (for $x \leq -\omega_b/2$ or $x \geq \omega_b/2$) where the center of the bar is defined as position zero. For ideal EPR entanglement shown in Exp. (1), the ghost imaging would show a pattern proportional to $|\Gamma_x|^2$ (Fig. 3(a)) while ghost interference experiment would reveal an interference pattern proportional to $|\bar{\Gamma}_{2\pi x/\lambda f_2}|^2$ (Fig. 3(b)), where superscript – represents Fourier Transformation.

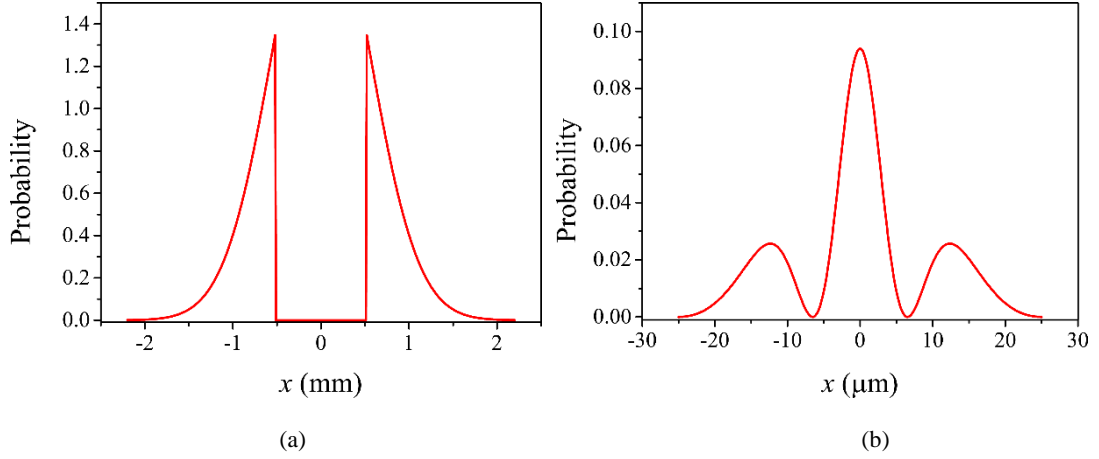

Fig 3. The normalized probability distributions of ghost imaging (a) and ghost interference (b) for ideal EPR entanglement

For experimental observation of ghost imaging/interference, we fit the data with the convolution of $|\Gamma_x|^2 / |\overline{\Gamma}_{2\pi x/\lambda f_2}|^2$ with a Gaussian function where we takes into account less-than-perfect EPR correlation and the finite resolution of collecting fiber. We use the same method in Ref. [10] to calculate the $(\Delta p_{x+})^2/(\Delta x_-)^2$ where the uncertainty is evaluated as the difference between the ideal curve and the resulting fitted curve for ghost interference/imaging.

At present, there are also some other methods to calculate the variances. For example, Kim's group directly fits single ghost interference/imaging through taking the phase matching condition and wave function in transverse position-momentum space together, thus getting the variances of positions and momenta simultaneously from only one figure [28]. And also Wasilewski's group fully takes the advantage of spatially resolving single-photon detectors to measure the wave function directly in position and momentum space [15].

## Author contributions

## Acknowledgements

This work was supported by National Key R&D Program of China (2017YFA0304800), the National Natural Science Foundation of China (Grant Nos. 61525504, 61722510, 61435011, 11174271, 61275115, 11604322), the Innovation Fund from Chinese Academy of Sciences, and Anhui Initiative In Quantum Information Technologies (AHY020200). We want to thank professor Jing Zhang (Shanxi University), professor Yoon-Ho Kim (Pohang University of Science and Technology) and professor Pei Wang (University of Science and Technology of China) for helpful discussions.

**References**

1. Einstein, A., Podolsky, B. & Rosen, N. Can Quantum-Mechanical Description of Physical Reality Be Considered Complete? *Phys. Rev.* **47**, 777–780 (1935).

2. Mancini, S., Giovannetti, V., Vitali, D. & Tombesi, P. Entangling Macroscopic Oscillators Exploiting Radiation Pressure. *Phys. Rev. Lett.* **88**, 120401 (2002).

3. Wagner, K. *et al*. Entangling the spatial properties of laser beams. *Science* **321**, 541–543 (2008)

4. Brida, G., Genovese, M. & Berchera, I. R. Experimental realization of sub-shot-noise quantum imaging. *Nat. Photon.* **4**, 227–230 (2010)

5. Pittman, T. B., Shih, Y. H., Strekalov, D. V. & Sergienko, A. V. Optical imaging by means of two-photon quantum entanglement. *Phys. Rev. A* **52**, R3429–R3432 (1995)

6. Tasca, D. S., Gomes, R. M., Toscano, F., Souto Ribeiro, P. H. & Walborn, S. P. Continuous-variable quantum computation with spatial degrees of freedom of photons. *Phys. Rev. A* **83**, 052325 (2011)

7. Reid, M. D. *et al*. The Einstein-Podolsky-Rosen paradox: From concepts to applications. *Rev. Mod. Phys.* **81**, 1727–1751 (2009)

8. Freedman, S. J. & Clauser, J. F. Experimental Test of Local Hidden-Variable Theories. *Phys. Rev. Lett.* **28**, 938–941 (1972).

9. Howell, J. C., Bennink, R. S., Bentley, S. J. & Boyd, R. W. Realization of the Einstein-Podolsky-Rosen Paradox Using Momentum and Position-Entangled Photons from Spontaneous Parametric Down Conversion. *Phys. Rev. Lett.* **92**, 210403 (2004)

10. D'Angelo, M., Kim, Y.-H., Kulik, S. P. & Shih, Y. H. Identifying Entanglement Using Quantum Ghost Interference and Imaging. *Phys. Rev. Lett.* **92**, 233601 (2004)

11. Moreau, P.-A., Devaux, F. & Lantz, E. Einstein-Podolsky-Rosen paradox in twin images. *Phys. Rev. Lett.* **113**, 160401 (2014)

12. Ou, Z. Y., Pereira, S. F., Kimble, H. J. & Peng, K. C. Realization of the Einstein-Podolsky-Rosen paradox for continuous variables. *Phys. Rev. Lett.* **68**, 3663–3666 (1992)

13. Ralph, T. C. & Lam, P. K. Teleportation with Bright Squeezed Light. *Phys. Rev. Lett.* **81**, 5668–5671 (1998)

14. Julsgaard, B., Kozhekin, A. & Polzik, E. S. Experimental long-lived entanglement of two macroscopic objects. *Nature* **413**, 400–403 (2001)

15. Dabrowski, M., Parniak, M. & and Wasilewski, W. Einstein-Podolsky-Rosen Paradox in a Hybrid Bipartite System. *Optica* **4**, 272–275 (2017)

16. Furusawa, A. *et al*. Unconditional quantum teleportation. *Science* **282**, 706–709 (1998)

17. Kuzmich, A. & Polzik, E. S. Atomic quantum state teleportation and swapping, *Phys. Rev. Lett.* **85**, 5639–5642 (2000)

18. Walborn, S. P., Lemelle, D. S., Almeida, M. P. & Souto Ribeiro, P. H. Quantum Key Distribution with Higher-Order Alphabets Using Spatially Encoded Qudits. *Phys. Rev. Lett.* **96**, 090501 (2006)

19. Edgar, M. P. *et al.* Imaging high-dimensional spatial entanglement with a camera. *Nat. Commun.* **3**, 984 (2012)

20. Braunstein, S. L. & van Loock, P. Quantum information with continuous variables. *Rev. Mod. Phys.* **77**, 513–577 (2005)

21. Duan, L. M., Lukin, M. D., Cirac, J. I. & Zoller, P. Long-distance quantum communication with atomic ensembles and linear optics. *Nature* **414**, 413–8 (2001)

22. Kimble, H. J. The quantum internet. *Nature* **453**, 1023–1030 (2008)

23. Jensen, K. *et al.* Quantum memory for entangled continuous-variable states, *Nat. Phys.* **7**, 13–16 (2011)

24. Marino, A. M., Pooser, R. C., Boyer, V. & Lett, P. D. Tunable delay of Einstein-Podolsky-Rosen entanglement. *Nature* **457**, 859–862 (2009)

25. Zhang, W. *et al*. Experimental realization of entanglement in multiple degrees of freedom

between two quantum memories. *Nature Communications* **7**, 13514 (2016)

26. Ding, D.-S., Zhang, W., Zhou, Z.-Y., Shi, S., Shi, B.-S. & Guo, G.-C. Raman quantum memory of photonic polarized entanglement. *Nat. Photon.* **9**, 332–338 (2015)

27. D'Angelo, M., Valencia, A., Rubin, M. H. & Shih, Y. H. Resolution of quantum and classical ghost imaging. *Phys. Rev. A* **72**, 013810 (2005)

28. Lee, J.-C., Park, K.-K., Zhao, T.-M. & Kim, Y.-H. Einstein–Podolsky–Rosen Entanglement of Narrow-band Photons from Cold Atoms. *Phys. Rev. Lett.* **117**, 250501 (2016)

29. Liu, Y. *et al.* Realization of a two-dimensional magneto-optical trap with a high optical depth. *Chin. Phys. Lett.* **29**, 024205 (2012)

30. Reid, M. D. Demonstration of the Einstein–Podolsky–Rosen paradox using nondegenerate parametric amplification. *Phys. Rev. A* **40**, 913–923 (1989)

31. Reid, M. D. *et al.* Colloquium: The Einstein–Podolsky–Rosen paradox: From concepts to applications. *Rev. Mod. Phys.* **81**, 1727 (2009).

32. Wiseman, H. M., Jones. S. J. & Doherty, A. C. Steering, Entanglement, Nonlocality, and the Einstein–Podolsky–Rosen Paradox. *Phys. Rev. Lett.* **98**, 140402 (2007)

33. Peise, J. *et al.* Satisfying the Einstein–Podolsky–Rosen criterion with massive particles. *Nature Communications* **6**, 8984 (2015)

34. Duan, L.-M., Giedke, G., Cirac, J. I. & Zoller, P. Inseparability Criterion for Continuous Variable Systems, *Phys. Rev. Lett.* **84**, 2722–2725 (2000)

35. Bennink, R. S., Bentley, S. J., Boyd, R. W. & Howell, J. C. Quantum and Classical Coincidence Imaging. *Phys. Rev. Lett.* **92**, 033601 (2004)

36. Ding, D.-S. *et al.* Quantum Storage of Orbital Angular Momentum Entanglement in an Atomic Ensemble. *Phys. Rev. Lett.* **114**, 050502 (2015).

37. Zhao, B. *et al*. A millisecond quantum memory for scalable quantum networks, *Nat. Phys.* **5**, 95–99 (2009).

38. Xu, Z.-X. *et al*. Long lifetime and high-fidelity quantum memory of photonic polarization qubit by lifting Zeeman degeneracy, *Phys. Rev. Lett.* **111**, 240503 (2013).

39. Radnaev, A. G. *et al*. A quantum memory with telecom-wavelength conversion, *Nat. Phys.* **6**, 894–899 (2010).

40. Yang, S.-J., Wang, X.-J., Bao, X.-H. & Pan, J.-W. An efficient quantum light–matter interface with sub-second lifetime. *Nat. Photon.* **10**, 381–384 (2016).

41. Heinze, G. *et al*. Stopped light and image storage by electromagnetically induced transparency up to the regime of one minute. *Phys. Rev. Lett.* **111**, 033601 (2013).

42. Chen, Y.-H. *et al*. Coherent Optical Memory with High Storage Efficiency and Large Fractional Delay. *Phys. Rev. Lett.* **110**, 083601 (2013).

43. Ding, D.-S. *et al*. High-dimensional entanglement between distant atomic-ensemble memories, *Light: Science & Applications* **5**, e16157(2016).

44. Ding, D.-S. Zhou, Z.-Y., Shi, B.-S. & Guo, G.-C. Single-photon-level quantum image memory based on cold atomic ensembles. *Nat. Commun.* **4**, 2527 (2013).

45. Nicolas, A. *et al.* A quantum memory for orbital angular momentum photonic qubits. *Nat. Photon.* **8**, 234–238 (2014).

46. O'Sullivan-Hale, M. N., Khan, I. A., Boyd, R. W. & Howell, J. C. Pixel Entanglement: Experimental Realization of Optically Entangled *d*=3 and *d*=6 Qudits. *Phys. Rev. Lett.* **94**, 220501 (2005)

47. Neves, L. *et al.* Generation of Entangled States of Qudits using Twin Photons. *Phys. Rev. Lett.* **94**, 100501 (2005)

48. Barreiro, J. T., Langford, N. K., Peters, N. A. & Kwiat, P. G. Generation of Hyperentangled Photon Pairs. *Phys. Rev. Lett.* **95**, 260501 (2005)